\documentclass[11pt, oneside]{amsart}

\usepackage{amsmath}
\usepackage{amssymb}

\usepackage{color}
\usepackage{dcolumn}
\usepackage{float}
\usepackage{graphicx}
\usepackage[latin9]{inputenc}
\usepackage{multirow}
\usepackage{rotating}
\usepackage{subfigure} 
\usepackage{psfrag}
\usepackage{tabularx}
\usepackage[hyphens]{url}
\usepackage{wrapfig}

\usepackage[bookmarks, bookmarksopen, bookmarksnumbered]{hyperref}
\usepackage[all]{hypcap}
\definecolor{orange}{rgb}{1.0,0.3,0.0}
\definecolor{violet}{rgb}{0.75,0,1}
\definecolor{darkgreen}{rgb}{0,0.6,0}
\definecolor{cyan}{rgb}{0.2,0.7,0.7}
\definecolor{blueish}{rgb}{0.2,0.2,0.8}

\usepackage[top=2.5cm, bottom=2.5cm, left=2.5cm, right=2.5cm]{geometry}

\sloppypar

\begin{document}

\title[]{Second Workshop on Sustainable Software for Science: Practice and Experiences (WSSSPE2): Submission, Peer-Review and Sorting Process, and Results}

\author{Daniel S. Katz$^{(1)}$, Gabrielle Allen$^{(2)}$, Neil Chue Hong$^{(3)}$, \\
Karen Cranston$^{(4)}$, Manish Parashar$^{(5)}$, David Proctor$^{(6)}$, \\
Matthew Turk$^{(2)}$, Colin C. Venters$^{(7)}$, Nancy Wilkins-Diehr$^{(8)}$}

\thanks{{}$^{(1)}$ National Science Foundation, Arlington, VA, USA; Computation Institute, University of Chicago \& Argonne National Laboratory, Chicago, IL, USA}

\thanks{{}$^{(2)}$ University of Illinois, Champaign, IL, USA}

\thanks{{}$^{(3)}$ Software Sustainability Institute, University of Edinburgh, Edinburgh, UK}
  
\thanks{{}$^{(4)}$ National Evolutionary Synthesis Center, Durham, NC, USA}

\thanks{{}$^{(5)}$ Rutgers Discovery Informatics Institute, Rutgers University, New Brunswick, NJ, USA}

\thanks{{}$^{(6)}$ International Consortium of Research Staff Associations, Dublin, Ireland}

\thanks{{}$^{(7)}$ University of Huddersfield, School of Computing and Engineering, Huddersfield, UK}

\thanks{{}$^{(8)}$ University of California-San Diego, San Diego, CA, USA}

\begin{abstract}
This technical report discusses the submission and peer-review process used by the Second Workshop on Sustainable Software for Science: Practice and Experiences (WSSSPE2) and the results of that process.  It is intended to record both the alternative submission and program organization model used by WSSSPE2 as well as the papers associated with the workshop that resulted from that process.
\end{abstract}

\maketitle

\section{Introduction}

%
%
%
%

The Second Workshop on on Sustainable Software for Science: Practice and Experiences (WSSSPE2)\footnote{\url{http://wssspe.researchcomputing.org.uk/wssspe2/}} will be held on Sunday, 16 November 2014, in conjunction with the 2014 International Conference for High Performance Computing, Networking, Storage and Analysis (SC14)\footnote{\url{http://sc14.supercomputing.org}}.
WSSSPE2 follows a general initial workshop, WSSSPE1\footnote{\url{http://wssspe.researchcomputing.org.uk/wssspe1/}}~\cite{WSSSPE1-pre-report,WSSSPE1}, that was held in 2013 in conjunction with the SC13 conference, and a focused workshop, WSSSPE1.1\footnote{\url{http://wssspe.researchcomputing.org.uk/wssspe1-1/}}, that was held in 2014 in conjunction with the SciPy conference.

Progress in scientific research is dependent on the quality and accessibility of software at all levels and it is critical to address challenges related to the development, deployment, maintenance and overall sustainability of reusable software as well as education around software practices. These challenges can be technological, policy based, organizational, and educational, and are of interest to developers (the software community), users (science disciplines), software engineering researchers, and researchers studying the conduct of science (science of team science, science of organizations, science of science and innovation policy, and social science communities). The WSSSPE1 workshop engaged the broad scientific community to identify challenges and best practices in areas of interest for sustainable scientific software. WSSSPE2 invites the community to propose and discuss specific mechanisms to move towards an imagined future practice for software development and usage in science and engineering. The workshop will include multiple mechanisms for participation, encourage team building around solutions, and identify risky solutions with potentially transformative outcomes. Participation by early career students and postdoctoral researchers is strongly encouraged.

\section{Submissions}

The workshop call for papers included the following areas of interest:
\begin{quote}
\begin{itemize}

\item defining software sustainability in the context of science and engineering software
\begin{itemize}
\item how to evaluate software sustainability
\end{itemize}

\item improving the development process that leads to new software
\begin{itemize}
\item methods to develop sustainable software from the outset
\item effective approaches to reusable software created as a by-product of research
\item impact of computer science research on the development of scientific software
\end{itemize}

\item recommendations for the support and maintenance of existing software
\begin{itemize}
\item software engineering best practices
\item governance, business, and sustainability models
\item the role of community software repositories, their operation and sustainability
\item reproducibility, transparency needs that may be unique to science
\end{itemize}

\item successful open source software implementations
\begin{itemize}
\item incentives for using and contributing to open source software
\item transitioning users into contributing developers
\end{itemize}

\item building large and engaged user communities
\begin{itemize}
\item developing strong advocates
\item measurement of usage and impact
\end{itemize}

\item encouraging industry's role in sustainability
\begin{itemize}
\item engagement of industry with volunteer communities
\item incentives for industry
\item incentives for community to contribute to industry-driven projects
\end{itemize}

\item recommending policy changes
\begin{itemize}
\item software credit, attribution, incentive, and reward
\item issues related to multiple organizations and multiple countries, such as intellectual property, licensing, etc.
\item mechanisms and venues for publishing software, and the role of publishers
\end{itemize}

\item improving education and training
\begin{itemize}
\item best practices for providing graduate students and postdoctoral researchers in domain communities with sufficient training in software development
\item novel uses of sustainable software in education (K-20)
\item case studies from students on issues around software development in the undergraduate or graduate curricula
\end{itemize}

\item careers and profession
\begin{itemize}
\item successful examples of career paths for developers
\item institutional changes to support sustainable software such as promotion and tenure metrics, job categories, etc.
\end{itemize}

\end{itemize}

\end{quote}

Based on the goal of encouraging a wide range of submissions from those involved in software practice, ranging from initial thoughts and partial studies to mature deployments, but focusing on papers that are intended to lead to changes, the organizers wanted to make submission as easy as possible.  The call for papers stated:

\begin{quote}
We invite short (4-page) \textbf{actionable} papers that will lead to improvements for sustainable software science. These papers could be a call to action, or could provide position or experience reports on sustainable software activities. The papers will be used by the organizing committee to design sessions that will be highly interactive and targeted towards facilitating action. Submitted papers should be archived by a third-party service that provides DOIs. We encourage submitters to license their papers under a Creative Commons license that encourages sharing and remixing, as we will combine ideas (with attribution) into the outcomes of the workshop.\end{quote}

31 submissions were received, and all but one used either arXiv\footnote{\url{http://arxiv.org}} or figshare\footnote{\url{http://figshare.com}} to self-publish their papers.  

\section{Peer-Review and Peer-Grouping}

The review process was fairly standard, where reviewers bid for papers, then an automated system
matched bids to determine assignments, and reviewers then completed their assigned
reviews (with an average of 4.9 reviews per paper, and 4.1 reviews per reviewer)
This process was done through EasyChair\footnote{http://easychair.org/}, which allowed reviewers to provide scores on relevance and comments to the organizers, which were used to decide which papers to associate with the workshop, and comments to the authors, which were provided back to the authors to allow them to improve their papers.

The organizers decided to list 28 of the papers as significantly contributing to the workshop, a very high acceptance rate, but one that is reasonable, given the goal of broad participation and the fact that the reports were already self-published.

WSSSPE1 was organized into sessions, each of which was aimed at discussing one or more of the themes from the call for papers, with a few paper authors invited to summarize the other papers in that them as a panel, followed by general discussion about that theme.  The mapping of papers to themes was done by the organizers.

For WSSSPE2, the organizers wanted to increase the interactivity of the sessions, and to open the process of creating the sessions to the full program committee, the paper authors, and others who might attend the workshop.  In order to do this, the organizers decided to use a breakout format for two sessions, and to use an open process to determine the breakout topics.  Specifically, well-sorted\footnote{\url{http://www.well-sorted.org}} was used as follows:
\begin{enumerate}
\item Authors were asked to create well-sorted ``cards'' for the papers.  These cards have a title (50 characters maximum) and a body (255 characters maximum).
\item Authors, program committee members, and members of the WSSSPE mailing list were asked to sort the cards.  Each person drags the cards, one by one, into groups.  A group can have as many cards as the person wants it to have, and it can have whatever meaning makes sense to that person.
\item Well-sorted then produces a set of averages of all the sorts, with various numbers of clusters of cards.
\end{enumerate}

The organizers then chose the sort that contained five groups as the one that felt most meaningful, then decided on names for the five groups, namely:
\begin{itemize}
\item Exploring Sustainability
\item Software Development Experiences
\item Credit \& Incentives
\item Reproducibility \& Reuse \& Sharing
\item Code Testing \& Code Review
\end{itemize}

Finally, since some of the papers were not represented by cards in the process, they were not placed in groups, so the authors of these papers were asked which group seemed the best for their papers, and those papers were then placed in those groups, as listed in the next section of this report.

\section{Results}

The contributed papers that will be discussed at the workshop follow, listed by groups (determined as described in the previous section.)

\begin{itemize}
\item Exploring Sustainability
\begin{itemize}
\item Mario {Rosado de Souza} and Robert Haines and Caroline Jay. Defining Sustainability through Developers' Eyes: Recommendations from an Interview Study~\cite{wssspe2_rosada_de_souza}
\item Robert Downs, W. Christopher Lenhardt, Erin Robinson, Ethan Davis, Nicholas Weber. Community Recommendations for Sustainable Scientific Software~\cite{wssspe2_downs}
\item Abani Patra, Matthew Jones, Steven Gallo, Kyle Marcus and Tevfik Kosar. Role of Online Platforms, Communications and Workflows in Developing Sustainable Software for Science Communities~\cite{wssspe2_patra}
\item Marlon Pierce, Suresh Marru and Chris Mattmann.{WSSSPE2}: Patching It Up, Pulling It Forward~\cite{wssspe2_pierce}
\item Justin Shi. Seeking the Principles of Sustainable Software Engineering~\cite{wssspe2_shi}
\item Colin C. Venters, Michael K. Griffiths, Violeta Holmes, Rupert R. Ward and David J. Cooke. The Nebuchadnezzar Effect: Dreaming of Sustainable Software through Sustainable Software Architectures~\cite{wssspe2_venters}
\end{itemize}

\item Software Development Experiences
\begin{itemize}
\item Jordan Adams, Sai Nudurupati, Nicole Gasparini, Daniel Hobley, Eric Hutton, Gregory Tucker and Erkan Istanbulluoglu. Landlab: Sustainable Software Development in Practice ~\cite{wssspe2_adams}
\item Alice Allen and Judy Schmidt. Looking before leaping: Creating a software registry~\cite{wssspe2_allen}
\item Carl Boettiger, Ted Hart, Scott Chamberlain and Karthik Ram. Building software, building community: lessons from the {ROpenSci} project~\cite{wssspe2_boettiger}
\item Michael R. Crusoe and C.Titus Brown. Channeling community contributions to scientific software: a hackathon experience~\cite{wssspe2_crusoe}
\item Yolanda Gil,  Eunyoung Moon and James Howison. No Science Software is an Island: Collaborative Software Development Needs in Geosciences~\cite{wssspe2_gil}
\item Ted Habermann, Andrew Collette, Steve Vincena, Werner Benger, Jay Jay Billings, Matt Gerring, Konrad Hinsen, Pierre de Buyl, Mark K\"{o}nnecke, Filipe Rnc Maia and Suren Byna. The Hierarchical Data Format ({HDF}): A Foundation for Sustainable Data and Software~\cite{wssspe2_habermann}
\item Marcus Hanwell, Patrick O'Leary and Bob O'Bara. Sustainable Software Ecosystems: Software Engineers, Domain Scientists, and Engineers Collaborating for Science~\cite{wssspe2_hanwell}
\item Eric Hutton, Mark Piper, Irina Overeem, Albert Kettner and James Syvitski. Building Sustainable Software - The {CSDMS} Approach~\cite{wssspe2_hutton}
\item W. Christopher Lenhardt, Stanley Ahalt, Matt Jones, J. Aukema, S. Hampton, S. R. Hespanh, R. Idaszak and M. Schildhauer. {ISEES-WSSI} Lessons for Sustainable Science Software from an Early Career Training Institute on Open Science Synthesis~\cite{wssspe2_lenhardt}
\item Jory Schossau and Greg Wilson. Which Sustainable Software Practices Do Scientists Find Most Useful?~\cite{wssspe2_schossau}
\item James S. Spencer, Nicholas S. Blunt, William A. Vigor, Fionn D. Malone, W. M. C. Foulkes, James J. Shepherd and Alex J. W. Thom. The {H}ighly {A}ccurate {N-DE}terminant ({HANDE}) quantum {Monte Carlo} project: Open-source stochastic diagonalisation for quantum chemistry~\cite{wssspe2_spencer}
\end{itemize}

\item Credit \& Incentives
\begin{itemize}
\item James Howison. Retract bit-rotten publications: Aligning incentives for sustaining scientific software~\cite{wssspe2_howison}
\item Daniel Katz and Arfon Smith. Implementing Transitive Credit with {JSON-LD}~\cite{wssspe2_katz}
\item Ian Kelley. Publish or perish: the credit deficit to making software and generating data~\cite{wssspe2_kelley}
\end{itemize}

\item Reproducibility \& Reuse \& Sharing
\begin{itemize}
\item Jakob Blomer, Dario Berzano, Predrag Buncic, Ioannis Charalampidis, Gerardo Ganis, George Lestaris and Ren\'{e} Meusel. The Need for a Versioned Data Analysis Software Environment~\cite{wssspe2_blomer}
\item Ryan Chamberlain and Jennifer Schommer. Using {Docker} to Support Reproducible Research~\cite{wssspe2_chamberlain}
\item Neil Chue Hong. Minimal information for reusable scientific software~\cite{wssspe2_chue_hong}
\item Tom Crick, Benjamin A. Hall and Samin Ishtiaq. ``Can I Implement Your Algorithm?'': A Model for Reproducible Research Software~\cite{wssspe2_crick}
\item Bryan Marker, Don Batory, Field G. Van Zee and Robert van de Geijn. Making Scientific Computing Libraries Forward Compatible~\cite{wssspe2_marker}
\item Stephen Piccolo. Building Portable Analytical Environments to improve sustainability of computational-analysis pipelines in the sciences~\cite{wssspe2_piccolo}
\end{itemize}

\item Code Testing \& Code Review
\begin{itemize}
\item Thomas Clune, Michael Rilee and Damian Rouson. Testing as an Essential Process for Developing and Maintaining Scientific Software~\cite{wssspe2_clune}
\item Marian Petre and Greg Wilson. Code Review For and By Scientists~\cite{wssspe2_petre}
\item Andrew E. Slaughter, Derek R. Gaston, John Peterson, Cody J. Permann, David Andrs and Jason M. Miller. Continuous Integration for Concurrent {MOOSE} Framework and Application Development on {GitHub}~\cite{wssspe2_slaughter}
\end{itemize}

\end{itemize}

\section{Other parts of the workshop}

WSSSPE2 will include two keynote presentations, and lightning talks from accepted paper authors, in addition to a set of breakout sessions.  The breakouts will allow attendees to participate in discussions about sustainability, future actions, and two of the five areas that resulted from the community submission, review, and grouping process.  There will also be reporting from the breakout groups to collect information for all attendees, as well as people who were not able to attend, to understand the discussions.
 
\section{Conclusions}

The WSSSPE2 workshop continues our experiment from WSSSPE1 in how we can collaboratively build a workshop agenda.  The differences in WSSSPE2 from WSSSPE1 are in using an existing service (EasyChair) to handle submissions and reviews, rather than an ad hoc process, and using an existing service (well-sorted) to allow collaborative grouping of papers into themes by all authors, reviewers, and the community, rather than this being done in an ad hoc manner by the organizers alone.

The fact remains that contributors also want to get credit for their participation in the process.  And the workshop organizers still want to make sure that the workshop content and their efforts are recorded.  Ideally, there would be a service that would be able to index the contributions to the workshop, serving the authors, the organizers, and the larger community.  But since there still isn't such a service today, the workshop organizers are writing this initial report and making use of arXiv as a partial solution to provide a record of the workshop.

After the workshop, one or more additional papers will be created that will include the discussions at the workshop.  These papers will likely have many authors, and may be submitted to peer-reviewed journals.

\bibliographystyle{plain}

\bibliography{wssspe}
\end{document}